\documentclass[11pt]{article}

\usepackage{graphics}

\begin{document}

\title{\bf K-CAUSAL STRUCTURE OF SPACE-TIME IN GENERAL RELATIVITY }
\date{}
\maketitle
$\ SUJATHA \ JANARDHAN^{1}$ and  $\ R \ V \ SARAYKAR^{2} $ \\
 $\ ^{1}$ Department of Mathematics, St.Francis De Sales
College,  Nagpur-440 006, India. \\ e-mail :
sujata\_jana@yahoo.com
\\ $\ ^{2}$ Department of Mathematics, R\ T\ M Nagpur University, Nagpur-440 033,
India. \\ e-mail :  r\_saraykar@rediffmail.com

\vspace{5mm} \textbf{Abstract.}   Using K-causal relation
introduced by Sorkin and Woolgar [26], we generalize results of
Garcia-Parrado and Senovilla [8 , 9] on causal maps. We also
introduce causality conditions with respect to K-causality which
are analogous to those in classical causality theory and prove
their inter relationships. We introduce a new causality condition
following the work of Bombelli and Noldus [3] and show that this
condition lies in between global hyperbolicity and causal
simplicity. This approach is simpler and more general as compared
to traditional causal approach [11 , 19] and it has been used by
Penrose et.al  [20] in giving a new proof of positivity of mass
theorem.$\ C^0 $ space-time structures arise in many mathematical
and physical situations like conical singularities, discontinuous
matter distributions, phenomena of topology-change in quantum
field theory etc.

\vspace{5mm} \textbf{Key  Words:} \ \  \ $\ C^{0}$- space
- times;\ causal maps;\ causality conditions;\ K-causality.

\vspace{5mm} \textbf{PACS Nos  \ \  \    04.20.-q; 02.40.Pc}

\vspace{8mm}

\section{Introduction}

\hspace{10mm} \ The condition that no material particle signals
can travel faster than the velocity of light fixes the causal
structure for Minkowski space-time. In general, causality is one
of the most important concepts in physical theories, and in
particular, in all relativistic theories based on a Lorentz
manifold. Thus in particular, in general relativity, locally the
causal relations are the same as in the Minkowski space-time.
However globally, there could be important differences in the
causal structure due to a change in space-time topology, or due to
strong gravitational fields etc. From physical point of view,
concept of causalities embodies the concept of time evolution,
finite speed of signal propagation and accessible communications.
Causality concepts were crucially used to formulate and prove the
singularity theorems [11, 14]. Several types of causality
conditions are usually required on space-times in order that these
space-times are physically reasonable.

It is also well - known that classical general relativity employs
a Lorentzian space - time metric whereas all reasonable approaches
to quantum gravity are free of such a metric background. Thus we
can ask the question whether there exists a structure which
captures essential features of light cones and hence notion of
causality in a purely topological and order theoretic manner
without a priori assumption of a Lorentzian metric. The results in
sections 2 and 3 in this paper show that this is possible. Thus by
using only topology and order defined on the cones, one can
recover different causality conditions, and also establish their
inter-relationships.

The order theoretic structures, namely causal sets, have been
extensively used by Sorkin and his co-workers in developing a new
approach to quantum gravity [23, 24, 25]. The use of domain
theory, a branch of computer science, by K. Martin [15] shows that
only order is necessary to define causal structures and the
topology is implicitly described by the order at an abstract
level. We shall comment more on this at the end of the paper.
Furthermore, the conformally invariant causal cone structures have
been used in resolving factor-ordering problems in the
quantization in mini superspace models of higher dimensional
Einstein gravity by Rainer[21, 22]. Thus we feel that the approach
and results in this paper will be helpful in studying such
phenomena and also phenomena like topology- change in quantum
gravity in a more general manner.

As far as$\ C^0 $- Lorentzian manifolds are concerned the use of
$\ C^0 $ structure does not need any differentiability condition
and thus maybe useful to analyze a wide variety of situations of
physical and mathematical interest, like conical singularities
(occurring in quantum field theory) and metric describing
discontinuous matter distributions.

Thus in section 2, using$\ C^0 $- Lorentzian manifold structure,
we define K-causal maps and prove their properties. Motivation for
this study is given in the beginning of section 2.

In section 3, we define a number of causality conditions with
respect to $\ K^{+} $ analogous to [11, 14, 19 ], and prove some
interesting properties of space-times satisfying these conditions
and their inter-relationships.

In section 4, we discuss the hierarchy of causality conditions on
the basis of results proved in section 3 and make some concluding
remarks.

\vspace{8mm}

\section{K-Causal Maps}

\hspace{10mm}\ In 1996, Sorkin and Woolgar [26] introduced a new
causal relation $\ K^{+}$ which is a generalization of
chronological relation $\ I^{+}$ and causal relation $\ J^{+}$ (
cf Penrose [19] and Hawking and Ellis [11]). Theory of causal
structures developed by Penrose and Hawking played an important
role in establishing singularity theorems in general relativity
and quite exhaustive work has been done in this area since 1964.
Books by Hawking and Ellis [11], Wald [27], Joshi [14] and Beem,
Ehlrich and Easley [2] are a proof of it. The causal relation $\
K^{+}$ introduced in [26] is order-theoretic in nature and no
smoothness assumption is needed to develop the ideas in General
Relativity. In Sorkin and Woolgar [26], the results are proved for
$\ C^{0}$-Lorentzian manifolds ( a $\ C^{1}$- manifold endowed
with a $\ C^{0}$- metric)and the framework developed is of wider
applicability as compared to previous work and also conceptually
simple. The authors have made use of Vietoris topology to
establish some basic results. An important result proved is the
compactness of the space of K-causal curves in a globally
hyperbolic space-time. More recently A. Garcia-Parrado and J.M.
Senovilla [8 , 9] introduced the concept of causal mappings and
proved a series of results in Causal structure theory of
space-times in General Relativity. These mappings are more general
than conformal mappings and the authors have also discussed a
number of examples to illustrate their ideas. Causal isomorphisms
used in [8 , 9] are a generalization of `chronal isomorphisms'
first used by Zeeman  in the context of Minkowski space and later
generalized by Joshi , Akolia and Vyas ( see Joshi [14] , sec 4.8
) for space-times in general relativity.

Dowker, Garcia and Surya [6] have proved that in a stably causal
space-time, K-causal future is equal to Seifert future. Recently,
Minguzzi [18] also proved some interesting results which may lead
to the proof of equivalence of K-causality and stable causality.We
shall comment more on this in the last section. Since we are
interested in the results not depending on metric, but only on
topology and order, we shall restrict to K-causal considerations.

In this  and the following section, we combine the concept of
K-causality with causal mappings and prove a series of results
which are generalizations of results of Garcia-Parrado and
Senovilla. In fact, our aim is to recast `global causal analysis'
along order-theoretic and general-topological lines, following the
definitions and basic results of Sorkin and Woolgar [26] and
following other works by Hawking and Sachs [12], Beem [1], Geroch
[10], Joshi [14] and Diekmann [5].

Thus we first recall basic definitions from Sorkin and Woolgar
[26] and then define K-causal maps and derive their properties. We
then discuss briefly the algebraic structure of the set of all
K-causal maps on a $\ C^{0}$- Lorentzian manifold. We also define
K-conformal maps and as expected, these maps preserve all K-causal
properties.

To begin with, we give  basic definitions and some results from
Sorkin and Woolgar [26] that will be used in this paper.
Throughout this paper, by a $\ C^{0}$- space-time V, we mean a $\
C^{0}$ -Lorentzian manifold with metric $\ g_{ab}$ ( a $\ C^{1}$-
manifold endowed with a $\ C^{0} $- metric).

\vspace{5mm}

\noindent\textbf{Definition (i) :}  Let V be a $\ C^{0}$- space -
time and $\ u^{a}$ be any vector field defining its time
orientation. A timelike or lightlike vector $\ v^{a}$ is
\emph{future-pointing} if $\ g_{ab}v^{a}u^{b} < 0 $ and
\emph{past-pointing} if $\ g_{ab}v^{a}u^{b} > 0 $. Now let $\ I =
[0,1]  $. A \emph{future-timelike path} in V is a piecewise $\
C^{1} $ continuous function $\ \gamma : [0,1] \rightarrow M $
whose tangent vector $\ \gamma^{a}(t) = (d \gamma(t)/dt)^{a}$ is
future- pointing timelike whenever it is defined.A
\emph{past-timelike} path is defined dually. The image of a
future- or past-timelike path is a \emph{timelike curve}. Let O be
an open subset of V . If there is a future-timelike curve in O
from p to q , we write $\ q \in I^{+} (p, O) $, and we call $\
I^{+}(p,O), $  the \emph{chronological future} of p relative to O
. Past-timelike paths and curves are defined analogously.

\vspace{5mm}

\noindent\textbf{Definition (ii):} \emph{$\ K^{+}$} is the
smallest relation containing $\ I^{+}$ that is topologically
closed  and transitive. If \emph{q is in $\ K^{+}(p)  $ then we
write $\ p \prec q $}.

 That is, we define the relation $\ K^{+}$ , regarded as a subset
of $\ V \times V $ , to be the intersection of all closed subsets
$\ R \supseteq I^{+} $ with the property that $\ (p, q)  \in  \ R
$ and $\ (q, r) \in \ R $   implies $\ (p, r) \in \ R $. ( Such
sets R exist because $\ V \times V $ is one of them.) One can also
describe $\ K^{+}$ as the closed-transitive relation generated by
$\ I^{+}$.

\vspace{5mm}

\noindent\textbf{Remark :} \ We note here that to define $\
K^{+}$, we need  $\ I^{+}$ and the topology of space-time
manifold. $\ I^{+}$ can be defined if, a priori, a cone structure
is given. Thus a cone structure and topology are sufficient to
define $\ K^{+}$.

\vspace{5mm}

\noindent\textbf{Definition (iii):}\ An open set O is
\emph{K-causal} iff the relation $\ \prec $ \ induces a reflexive
partial ordering on O. i.e., $\ p \prec q $ and $\ q \prec p $
together imply $\ p = q $.

\vspace{5mm}

\noindent\textbf{Definition (iv):}\ A subset of V is
\emph{K-convex} iff it  contains along with p and q any $\ r\in V
$ for which p $\ \prec $ r$\ \prec $ q.

\vspace{5mm}

\noindent\textbf{Definition (v):}\ A \emph{K-causal curve} $
\Gamma $ from p to q is the image of a $\ C^{0} $  map  $\ \gamma:
[0,1]\rightarrow V $ with $\ p=\gamma(0), q=\gamma(1) $ and such
that for each $\ t\in (0,1) $ and  each open set $\ O \ni
\gamma(t)$ there is a positive number $\ \epsilon $ such that \\
$\ t^{'}\in(t,t+\epsilon) \Rightarrow \gamma(t) \prec_{O}
\gamma(t^{'}).\\t^{'}\in(t-\epsilon,t) \Rightarrow \gamma(t^{'})
\prec_{O} \gamma(t^)$.

Here $\ \prec_{O} $ denotes $\ K^{+} $ relation relative to O.

\vspace{5mm}

\noindent\textbf{Remark :}\ In the standard formalism, a causal
curve has to satisfy a differentiability condition because one has
to define its tangent vector in order to know if the tangent
vector lies within the light cone or not. The differentiability
condition differs from one author to another. For example, Penrose
[19] requires time like curves to be everywhere differentiable and
smooth, however, he works mostly with \emph{trips}, which are only
piecewise - smooth. We know that, not all continuous curves are
differentiable, smooth, or even piecewise smooth. Indeed, there
are continuous curves which fail to be differentiable at
infinitely many points. If such a curve can be linearly ordered by
K, then it is a K-causal curve but not a causal curve.

\vspace{5mm}

\noindent\textbf{Definition (vi):}\ A K-causal open set $\ O
\subseteq V $ is \emph{globally hyperbolic }
 iff for every pair of points $\ p ,q \in O,$ the interval $\ K(
 p,q ) = K^{+}(p) \cap \ K^{-}(q) $ is compact and contained in $\ O $.

\vspace{5mm}

We shall need the following theorems which are proved in Sorkin
and Woolgar [26]:

\vspace{5mm}

\noindent\textbf{Theorem (i):}\ If V is K-causal then every
element of V possesses an arbitrarily small K-convex open
neighbourhood.

\vspace{5mm}

\noindent\textbf{Theorem (ii):}\ A subset $\ \Gamma $ of a
K-causal space - time is a K-causal curve iff it is
compact,connected and linearly ordered by $\ \prec  = K^{+} $.

\vspace{5mm}

\noindent\textbf{Theorem (iii):} \ In a K-causal space - time, let
the K-causal curve $\ \Gamma $ be the Vietoris limit of a sequence
of K-causal curves $\ \Gamma_{n}$ with initial endpoints $\ p_{n}$
and final endpoints $\ q_{n} $. Then $\ p_{n}$ converge to the
initial endpoint of $\ \Gamma $ and $\ q_{n} $ to its final
endpoint.

\vspace{5mm}

\noindent\textbf{Theorem (iv):} In a K-causal space - time , the
Vietoris limit of a sequence (or net) of causal curves is also a
causal curve.

\vspace{5mm}

\noindent\textbf{Theorem (v):} \  Let $\ O $ be a globally
hyperbolic open subset of a space - time V  and let $\ P $ and $\
Q $ be compact subsets of $\ O $. Then the space of  K-causal
curves from $\ P $ to $\ Q $ is bicompact.

\vspace{5mm}

With this background we now define a K-causal map. We work with $\
K^{+} $ throughout. Analogous definitions and results for $\ K^{-}
$ can be derived similarly.

\vspace{5mm}

A K- causal map is a causal relation which is a homeomorphism
between the two topological spaces and at the same time preserves
the order with respect to $\ K^{+}$. To define it, we first define
an order preserving map with respect to $\ K^{+} $:

\vspace{5mm}

\noindent\textbf{Definition 1:} Let V and W be $\ C^{0}$- space -
times. A mapping $\ f: V \rightarrow W $ is said to be \emph{order
preserving with respect to} $\ K^{+} $ or simply \emph{order
preserving }if whenever $\ p,q \in V $ with $\ q \in K^{+}(p),$ we
have $\ f(q) \in K^{+}(f(p))$. i.e., $\ p \prec q $ implies $\
f(p) \prec \ f(q) $.

\vspace{5mm}

\noindent\textbf{Definition 2:}\ Let V and W be $\ C^{0}$- space -
times. A homeomorphism $\ f: V \rightarrow W $ is said to be
\emph{K-causal} if $\ f $ is order preserving.

\vspace{5mm}

\noindent\textbf{Remark :}\ In general K-causal maps and causal
maps defined by A. Garcia-Parrado and J.M. Senovilla [8 , 9] are
not comparable as $\ r \in K^{+}(p) $ need not imply that $\ r \in
I^{+}(p) $ (cf figure 1).

\vspace{5mm}

Using the definition of K-causal map, we now prove a series of
properties which follow directly from the definition. We give
their proofs for the sake of completeness:

\vspace{2mm}

\noindent \emph{`A homeomorphism $\ f: V \rightarrow W $  is order
preserving iff
\\ $\ f((K^{+}(x))\subseteq K^{+}(f(x)), \forall \ x \in V '$.
}This is proved as below:

Let $\ f: V \rightarrow W $ be an order preserving homeomorphism
and let $\ x \in V $. Let $\ y \in f(K^{+}(x))$. Then $\ y = f(p),
\  x \prec p $ which implies $\ f(x) \prec \ f(p)$ as $\ f $is
order preserving. i.e., $\
f(x)\prec y $ or  $\ y \in \ K^{+}(f(x))$. \\
Hence $\ f(K^{+}(x)) \subseteq \ K^{+}(f(x)), \ \forall \ x \ \in \ V $.\\
Conversely let $\ f: V \rightarrow W $ be a homeomorphism such
that \\ $\ f((K^{+}(x)) \subseteq  \ K^{+}(f(x)), \ x \in \  V $. \\
Let $\ p \prec q $. Then  $\  f(q) \in \ f(K^{+}(p))$. By
hypothesis, this gives  $\ f(q) \in \ K^{+}(f(p))$. Hence $\ f(p)
\prec \ f(q)$. Thus if $\ f $ is a K-causal map then $\
f(K^{+}(x)) \subseteq \ K^{+}(f(x)),  \forall \ x \in V $.

\vspace{5mm}

Similarly we have the property that, \emph{`If $\ f: V \rightarrow
W $ be a homeomorphism then $\ f^{-1}$ is order preserving iff $\
K^{+}(f(x))\subseteq f((K^{+}(x)), \ x \in V $'.}

\vspace{5mm}

We now define, for $\ S\subseteq V,  $   $\  K^{+}(S)=  \bigcup_{x
\in S } K^{+}(x)  $.

\noindent In general, $\ K^{+}(S) $ is neither open nor closed. In
section 3, we shall show that in a globally hyperbolic $\ C^{0} $
- space - time, if S is compact, then $\ K^{+}(S) $ is closed.
However at present, we can prove the following property that

\noindent \emph{`If $\ f: V \rightarrow W $  is an order
preserving homeomorphism then $\ f((K^{+}(S))\subseteq
K^{+}(f(S)), \ S\subseteq  V $'.}

For, if $\ f: V \rightarrow W $ be an order preserving
homeomorphism and \\ $\ S\subseteq V $ then by definition, $\
K^{+}(S)= \bigcup_{x\in S} K^{+}(x)$. Let $\ y \in f(K^{+}(S)) $.
Then there exists $\ x $ in S such that $\ y
 \in f(K^{+}(x))$. This gives $\ y \in K^{+}(f(x))$.\\
 i.e., $\ y \in K^{+}(f(S))
$. Hence $\ f(K^{+}(S)) \subseteq  K^{+}(f(S))$.

\vspace{2mm}

Analogously we have, \emph{`If $\ f: V \rightarrow W $ be a
homeomorphism, and $\ f^{-1}$ is order preserving then $\
K^{+}(f(S)) \subseteq \ f(K^{+}(S)), \ S \subseteq \ V $'.}

\vspace{5mm}

We know that causal structure of space-times is given by its
conformal structure. Thus, two space-times have identical
causality properties if they are related by a conformal
diffeomorphism. Analogously, we expect that a K- conformal map
should preserve K- causal properties. Thus we define a K-
conformal map as follows.

\vspace{5mm}

\noindent\textbf{Definition 3:}\ A homeomorphism $\ f: V
\longrightarrow W $ is said be K-conformal if both $\ f $ and $\
f^{-1} $ are K-causal maps.

\vspace{5mm}

\noindent\textbf{Remark :}\ A K-conformal map is a causal
automorphism in the sense of E.C.Zeeman [28].

\vspace{5mm}

\noindent This definition is similar to \emph{chronal / causal
isomorphism} of Zeeman [28], Joshi [14] and Garcia - Parrado and
Senovilla [8 , 9].

\vspace{5mm}

Combining the above properties , we have the following,
namely,\emph{ If $\ f: V \rightarrow \ W $ is K-conformal then $\
f(K^{+}(x)) = K^{+}(f(x)), \forall \ x \in V $.}

\vspace{5mm}

By definition, K- conformal map will preserve different K-
causality conditions defined in section 3, below. If a map is only
K- causal and not K- conformal, then we have the following
properties:

\noindent To begin with, \emph{`If  $\ f: V \rightarrow W $ is a
K-causal mapping and W is K-causal, so is V'.}

For, let $\ f: V \rightarrow W $ be a K-causal map and W be
K-causal. Let $\ p \prec q $ and  $\  q \prec p, \ p, \ q \in V $.
Then $\ f(p), \ f(q) \ \in W $ such that $\ f(p) \prec f(q)$ and
$\ f(q) \prec f(p)$ as $\ f $ is order preserving. Therefore $\
f(p) = f(q) $ since W is K-causal. Hence $ \ p = q $.

\vspace{2mm}

Analogous result would follow for $\ f^{-1} $. In addition, a
K-causal mapping takes K-causal curves to K-causal curves.This is
given by the property that,  \emph{`If V be a K-causal space -
time and $\ f:V \rightarrow W $ be a K-causal mapping, then $\ f $
maps every K-causal curve in V to a K-causal curve in W'}, which
is proved as below:

Let $\ f: V \rightarrow W $  be a K-causal map. Therefore $\ f $
is an order preserving homeomorphism. Let $\ \Gamma $ be a
K-causal curve in V. Then by theorem (ii), $\ \Gamma $ is
connected, compact and linearly ordered. Since $\ f $ is
continuous, it maps a connected set to a connected set and a
compact set to a compact set. Since $\ f $ is order preserving and
$\ \Gamma $ is linearly ordered, $\ f(\Gamma) $ is a K-causal
curve in W. Analogous result would follow for $\ f^{-1} $.

\vspace{5mm}

From the above result we can deduce that, \emph{`If $\ f$ be a
K-causal map from V to W, then for every future directed K-causal
curve $\ \Gamma $ in V, any two points $\ x, \ y \in \ f(\Gamma) $
satisfy $\ x \prec y \ or \ y \prec x $'.}

\vspace{5mm}

\noindent\textbf{Definition 4:}\ Let V and W be two $\ C^{0}$ -
space - times. Then W is said to be \emph{K-causally related }to V
if there exists a K-causal mapping $\ f $ from V to W. i.e., $\ V
\prec_{f}W $.

\vspace{5mm}

\noindent The following property follows easily from this
definition, which shows that the relation $\ `\prec_{f}'$ is
transitive also.

\noindent \emph{If $\ V \prec_{f}W $ and $\ W \prec_{g}U $ then $\
V \prec_{g \circ f}U.$}

\vspace{5mm}

\noindent Next, we have the  property concerning K-convex sets,
that \emph{` If $\ f: V \longrightarrow W $ is a K-causal map then
$\ C \subseteq V $ is K-convex if $\ f(C)$ is a K-convex subset of
W'}. The proof is as below:

Let $\ f: V \longrightarrow W $ be K-causal and $\ f(C) $ be a
K-convex subset of W. Let $\ p , q \in C $ and $\ r \in V $ such
that $\ p \prec r \prec q $. Since $\ f $ is order preserving we
get  $\ f(p) \prec f(r) \prec f(q) $ where $\ f(p),f(q) \in f(C) $
and $\ f(r) \in W $. Since $\ f(C) $ is K-convex, $\ f(r) \in f(C)
$. i.e., $\ r \in C $. Hence C is a K-convex subset of V.

\vspace{5mm}

\noindent\textbf{Remark:}\ Concept of a convex set is needed to
define `strong causality', as we shall see below.

\vspace{5mm}

We now discuss briefly the algebraic structure of the set of all
K-causal maps from V to V. We define the following:

\vspace{5mm}

\noindent\textbf{Definition 5:} \ If V is a $\ C^{0}$ - space -
time then \emph{Hom(V)} is defined as the group consisting of all
homeomorphisms acting on V.

\vspace{5mm}

\noindent\textbf{Definition 6:} \ If V is a $\ C^{0}$ - space -
time then \emph{K(V)} is defined as the set of all K-causal maps
from V to V.

\vspace{5mm}

\noindent Then we have the property that \emph{`K(V) is a
submonoid of Hom(V)'}, which is more or less obvious.

\noindent \ For, if $\ f_{1}, f_{2}, f_{3} \in K(V) $ then  $\
f_{1} \circ f_{2} \in K(V)$. Also,  $\ f_{1} \circ ( f_{2} \circ
f_{3}) = (f_{1} \circ  f_{2})
 \circ f_{3}$  and identity homeomorphism exists. Hence K(V) is a
submonoid of Hom(V).

\vspace{5mm}

It is obvious that K(V) is a bigger class than the class of
K-conformal maps.

\vspace{8mm}

\section{K-causal structure}

\hspace{5mm} \ Different causality conditions are imposed on a
space-time to make it a physically reasonable one. In other words,
this is to avoid different pathological situations which would
arise otherwise.

In this section, we define different causality conditions with
respect to K-causality analogous to those in causal structure
theory and prove their properties, and inter-relationship. We also
illustrate with the help of diagrams that these conditions are not
equivalent to those in standard causal structure theory. We
introduce a condition coined by Noldus and interestingly enough,
we show that it is intermediate between global hyperbolicity and
causal simplicity. Finally, we get a simpler hierarchy among K-
causality conditions. Needless to say that K- conformal maps will
preserve all these K- causality conditions.

To begin with, analogous to usual causal structure, we define
strongly causal and future distinguishing space-times with respect
to $\ K^{+}$ relation.

\vspace{5mm}

\noindent\textbf{Definition 7:} \  A $\ C^{0} $ - space - time V
is said to be \emph{strongly causal at p with respect to $\
K^{+}$}, if  p has arbitrarily small K - convex open
neighbourhoods.

V is said to be \emph{strongly causal with respect to $\ K^{+}$},
if it is strongly causal  at each and every point of it with
respect to $\ K^{+}$. Thus, lemma 16 of [26] implies that
K-causality implies strong causality with respect to $\ K^{+} $.
Analogous definition would follow for $\ K^{-} $.

\vspace{5mm}

\noindent\textbf{Definition 8:} \ A $\ C^{0} $- space - time V is
said to be \emph{K-future distinguishing} if for every $\ p \neq q
, K^{+}(p) \neq K^{+}(q) $. \ \emph{K-past distinguishing} spaces
can be defined analogously.

\vspace{5mm}

\noindent\textbf{Definition 9:} \ A $\ C^{0} $- space - time V is
said to be \emph{K-distinguishing }if it is both K-future and
K-past distinguishing.

\vspace{5mm}

\noindent Since,  for every $\ p \neq q , I^{+}(p) =  I^{+}(q)$
implies $\ K^{+}(p) =  K^{+}(q) $, we have the following property:

\emph{If a $\ C^{0}$- space - time V is K- future distinguishing
then it is future distinguishing.}

\vspace{5mm}

\noindent\textbf{Remark :} \ In general, future distinguishing
need not imply K-future distinguishing.(refer  figure 3). Figure 2
shows a space-time which is neither future distinguishing nor
K-future distinguishing. However, we have the following the
following property:
\newline \ \emph{If a $\ C^{0}$- space - time V is K-causal, then it is
K-distinguishing}. The proof is as follows:

Let $\ p , q \in V $ be such that, $\ K^{+}(p) =  K^{+}(q) $. Now
$\ p \in \ K^{+}(p) $ implies $\ p \in K^{+}(q) $. Similarly, $\ q
\in K^{+}(p) $. Thus, by K-causality, $\ p \ = \ q $. Hence, V is
K-future distinguishing. Similarly, V is K-past distinguishing
also. Thus, every K-causal space-time is K-distinguishing.

\vspace{2mm}

We have a stronger result in the property, namely ,\emph{`If a $\
C^{0}$- space - time V is strongly causal with respect to $\ K^{+}
$, then it is K-future distinguishing'}, the proof of which is as
follows:

Let V be strongly causal with respect to $\ K^{+} $. Let $\ p, q
\in V $ such that $\ p  \neq q $ and $\ K^{+}(p) = K^{+}(q) $.

Let P and Q be disjoint K-convex neighbourhoods of p and q in V
respectively . Choose $\ x \in K^{+}(p) \cap P $. Then $\ x \in
K^{+}(p)$. Therefore, $\  x \in K^{+}(q)$. Let $\ y$ in Q be such
that $\ q \prec y \prec x $. Then $\ y$ is not in P and $\ y \in
K^{+}(q) $ which implies  $\ y \in K^{+}(p)$. So, $\  p \prec y
\prec x $, but y not in P, contradicting the hypothesis that P is
K- convex. Hence, $\ K^{+}(p) \neq \
 K^{+}(q) $ whenever $\ p \neq \ q $.

\vspace{5mm}

Analogous result would follow for $\ K^{-} $. \emph{Hence, in  a
$\ C^{0}$- space - time V ,  strong causality with respect to K
implies K-distinguishing.}

\vspace{5mm}

\noindent\textbf{Remark  :} \ K-conformal maps preserve K-
distinguishing,  strongly causal with respect to $\ K^{+}$ and
globally hyperbolic properties.

\begin{figure}[c]

\begin{center}

\includegraphics{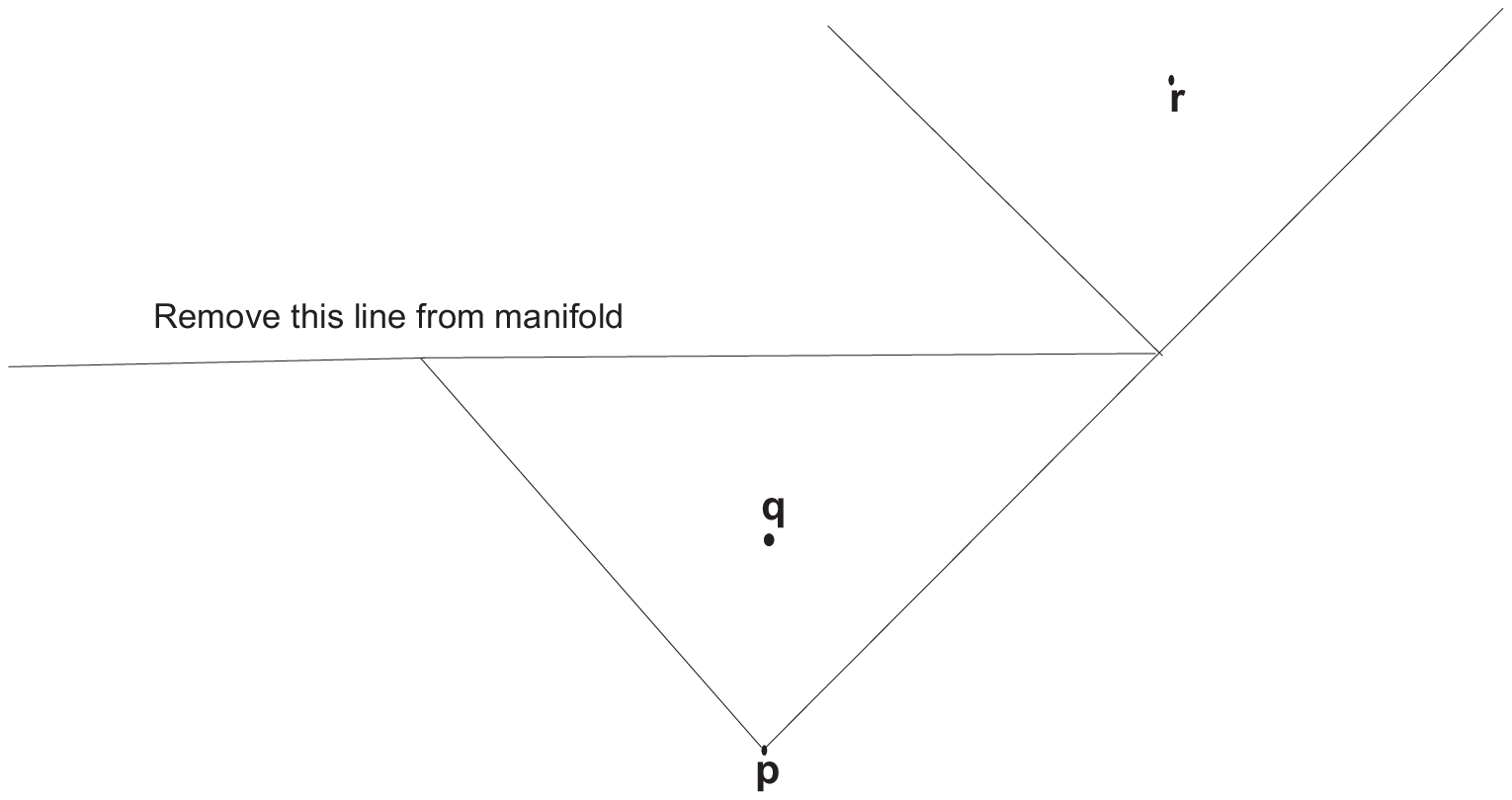}

\caption{ Two dimensional Minkowski space,  where  a closed
half-line is removed. Here, $\ r $ is in $\ K^{+}(p) $ but not in
$\ I^{+}(p)$. Also, $\ cl( I^{+}(p))  \neq \ K^{+}(p) $.}

\end{center}

\end{figure}

\begin{figure}[c]

\begin{center}

\includegraphics{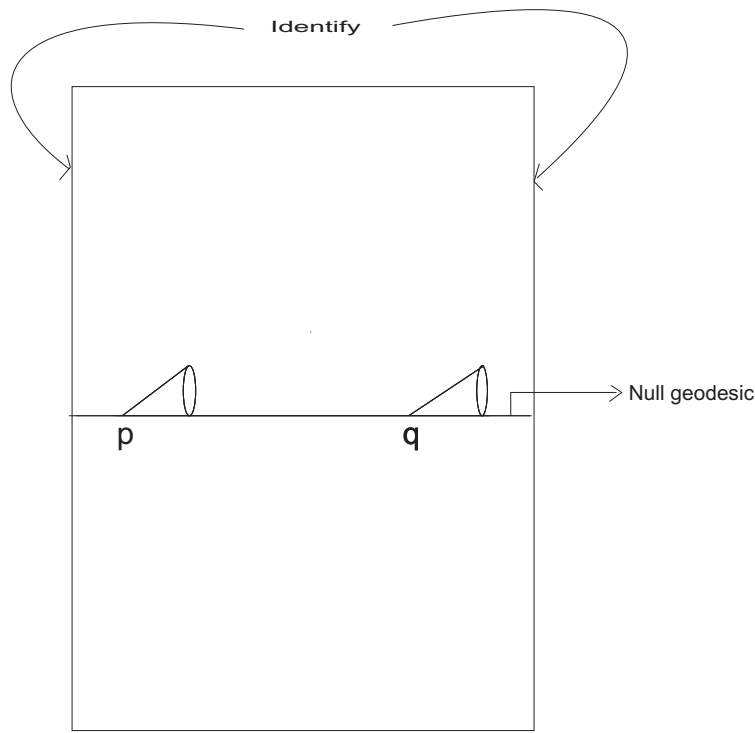}

\caption{A space-time which is neither future distinguishing nor
K-future distinguishing. Here, $\ p \neq \ q $ , but $\ I^{+}(p) =
\ I^{+}(q) $ and $\ K^{+}(p) = \ K^{+}(q) $ }

\end{center}

\end{figure}

\begin{figure}[c]

\begin{center}

\includegraphics{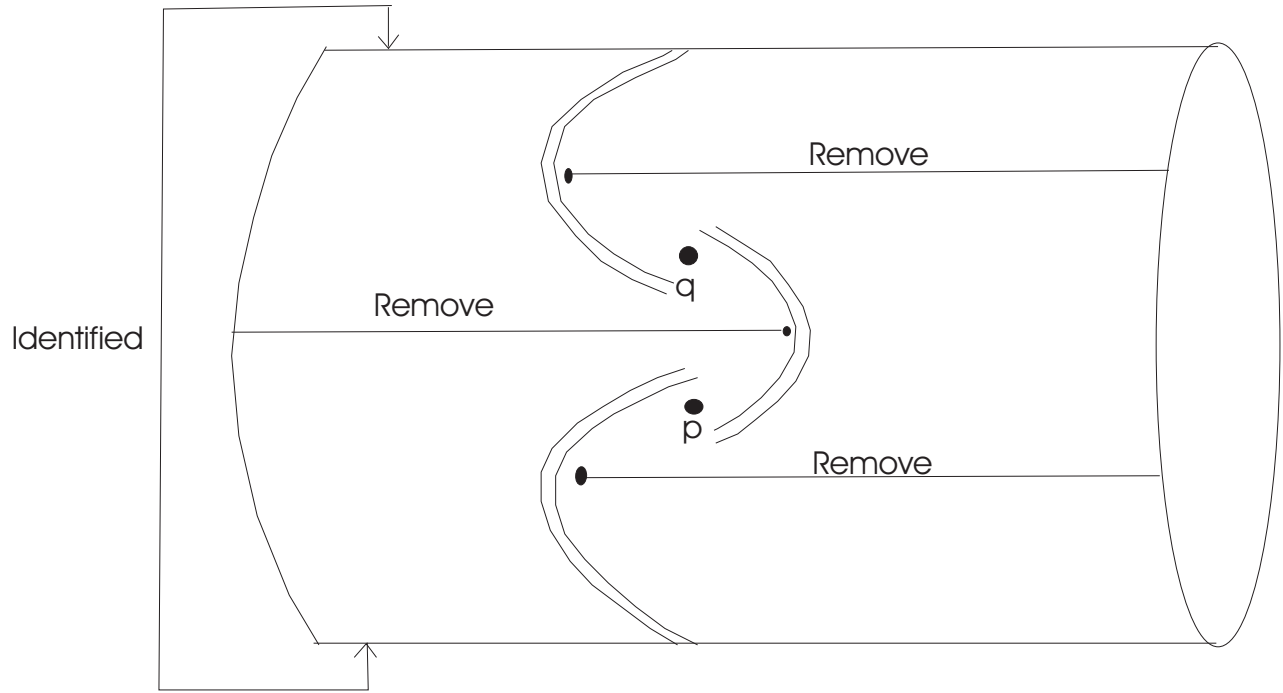}

\caption{ A time-like cylinder derived from $\ M^{2} $ with
Cartesian coordinates (t , z) by periodically identifying t .From
this space-time , remove three closed half -lines , each parallel
to the z -axis: the half -lines starting at ($\ \pm 1,-1 $) and
moving off to the right , and the half -line starting from (0, 0)
and moving off to the left . The resulting manifold has families
of time-like curves as depicted , which implies that $\ p \neq \ q
, I^{+} (p) \neq \ I^{+} (q), $ but $\ K^{+} (p) = \ K^{+} (q) $.
This is because $\ q \in \ K^{+}(p) $, and because of
identification, $\ p \in \ K^{+}(q)$. Hence , the space-time is
future distinguishing but not K-future distinguishing .}

\end{center}

\end{figure}

\begin{figure}[c]

\begin{center}

\includegraphics{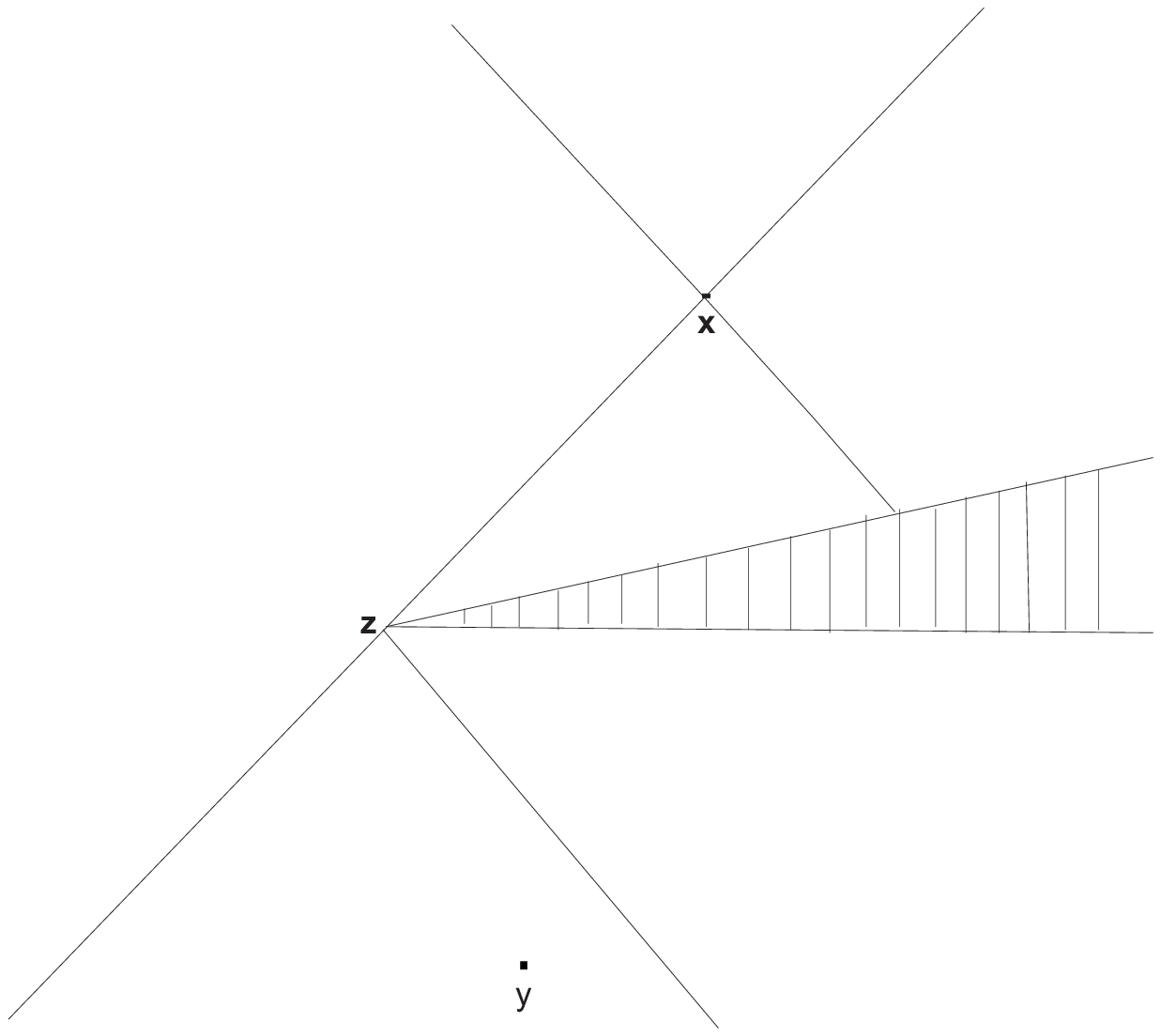}

\caption{ The space-time which is K-reflecting but not reflecting.
The shaded region represents closed subset which has been removed.
Here $\ z $ is in $\ K^{-}(x) $ but   not in $\ I^{-}(x)$. Hence,
$\ K^{+}(x) \subseteq K^{+}(y) \ \Rightarrow
 \ K^{-}(y) \subseteq K^{-}(x) $ and $\ I^{+}(x) \subseteq I^{+}(y)$
but $\ I^{-}(y) $ is not contained in $\ I^{-}(x) $. As any
K-causal space-time is K-reflecting, any non-reflecting open
subset of the space - time  will be K-causal but non-reflecting.}

\end{center}

\end{figure}

\vspace{5mm}

A reflecting condition is a useful causality condition and as
Clarke and Joshi [4] have proved, any stationary space-time is
reflecting. In our case, we may define K-reflecting space-times as
follows:

\vspace{5mm}

\noindent\textbf{Definition 10:} \ A $\ C^{0} $- space - time V is
said to be \emph{K-reflecting} if  \\ $\ K^{+}(p) \supseteq
K^{+}(q)  \Leftrightarrow \  K^{-}(q) \supseteq K^{-}(p) $.

\vspace{5mm}

However, since the condition $\ K^{+}(p) \supseteq  K^{+}(q) $
always implies $\ K^{-}(q) \supseteq K^{-}(p) $ because of
transitivity and $\ x \in K^{+}(x) $, and vice versa, a $\ C^{0}
$- space - time with K-causal condition is always K-reflecting. We
shall comment more on this in the concluding section.

\vspace{5mm}

Moreover, in general, K-reflecting need not imply reflecting
(refer figure 4). Since, any K-causal space-time is K-reflecting,
any non-reflecting open subset of the space - time will be
K-causal but non-reflecting.

\vspace{5mm}

Our last definition in this section is that of null relation with
respect to $\ K^{+} $. We define it as follows:

\vspace{5mm}

\noindent\textbf{Definition 11:} \ In V,  $\ y$ is said to be
\emph{ null related to x  with respect to $\ K^{+}$ }, if $\
K^{+}(x) \cap K^{-}(y) $ does not contain any open set. ( In this
case, we can say that $\ x $ and $\ y $ are null-related to each
other.)

\vspace{2mm}

\noindent Then we have the following property: \newline \
\emph{`If $\ f^{-1} : W \rightarrow V $ is a K-causal map, then $\
f $ preserves the null relation with respect to $\ K^{+} $'},
which can be proved as follows:

Let $\ x $ and $\ y$  be  null related  with respect to $\ K^{+}$.
Then, $\ K^{+}(x) \cap K^{-}(y) $ does not contain any open set.
Since $\ f $ is a homeomorphism,$\
 f( K^{+}(x) \cap K^{-}(y) )$ does not contain any open set. i.e.,
 $\ f( K^{+}(x)) \cap f( K^{-}(y)) $ does not contain any open set.
 Since, $\ f^{-1} $ is K-causal,  $\ K^{+}(f(x)) \cap K^{-}(f(y)) $
 does not contain any open set. Thus $\ f(x) $ and $\ f(y) $ are  null
related  with respect to $\ K^{+} $.

Similarly, if $\ f  $ is a K-causal map, then $\ f ^{-1} $
preserves the null relation with respect to $\ K^{+} $.

Hence, \emph{if $\ f $ is a K- conformal map, then $\ x $ and $\ y $ are
null-related iff $\
 f(x)$ and $\ f(y) $ are null-related with respect to $\ K^{+} $}.
Analogous result can be proved for $\ K^{-} $.

\vspace{5mm}

Since global hyperbolicity is the strongest causality condition
under study, we expect that a globally hyperbolic space-time with
respect to $\ K^{+} $ is a nicely behaving one, that is, there are
no pathologies in such a space-time. Thus we  expect that in such
a space-time, $\ K^{+}(x) $ = closure of $\ I^{+}(x) $, $\
I^{+}(x) $ = interior of $\ K^{+}(x) $, and thus boundaries of $\
K^{+}(x) $ and $\ I^{+}(x) $ coincide. We prove below that it is
indeed the case.

Also, in general $\ K^{+}(S)$ is not a closed set. We show that in
a globally hyperbolic space-time, if S is compact, then $\
K^{+}(S)$ comes out to be a closed set. We prove this as our first
result below:

\vspace{5mm}

\noindent\textbf{Theorem 1:} \ Let V be a globally hyperbolic $\
C^{0}$- space - time. If $\ S \subseteq V $ is compact then $\
K^{+}(S) $ is closed.

\noindent\textbf{Proof :} \ Let $\ S \subseteq V $ be compact. Let
$\ q \in \ cl(K^{+}(S))$. Then there exists a sequence $\ q_{n}$
in $\ K^{+}(S) $ such that $\ q_{n}$ converges to q. Hence there
exists a sequence $\ p_{n}$ in S  corresponding to $\ q_{n}$ and
future directed K- causal curves $\ \Gamma_{n}$ from $\ p_{n}$ to
$\ q_{n}$. Then $\ p_{n}$ has a subsequence $\ p_{n_{k}}$
converging to $\ p\in S $ since S is compact, which gives a
subsequence $\ \Gamma_{n_{k}}$ of future directed K-causal curves
from $\ p_{n_{k}}$ to $\ q_{n_{k}}$ where $\ p_{n_{k}}$ converges
to p and $\ q_{n_{k}}$ converges to q.

Define P = $\  \{p_{n_{k}} , p\} $ and Q = $\  \{q_{n_{k}} , q \}
$. Then P and Q are compact subsets of V. Hence, by theorem: iv,
 the set \emph{\textbf{C}} of all future
 directed K-causal curves from P to Q is compact. Now, $\ \{ \Gamma_{n_{k}}\}  $
 is a subset of \emph{\textbf{C}}. Thus, $\ \{ \Gamma_{n_{k}} \}
$ is a sequence in a compact set and hence has a convergent
subsequence say $\ \Gamma_{n_{k_{l}}} $ of future directed
K-causal curves from $\ p_{n_{k_{l}}} $ to $\ q_{n_{k_{l}}} $
where $\ p_{n_{k_{l}}} $ converges to p and  $\ q_{n_{k_{l}}} $
converges to q .Let $\ \Gamma $ be the Vietoris limit of $\
\Gamma_{n_{k_{l}}} $. Then by theorem (iii), $\ \Gamma $ is a
future directed K-causal curve from p to q . Since $\ p \in S $,
we have, $\ q\in K^{+}(S)$. Hence $\ cl(K^{+}(S)) \subseteq
K^{+}(S)$. Thus $\ K^{+}(S) $ is closed.

\vspace{5mm}

The next two theorems show that in a globally hyperbolic $\ C^{0}
$ - space - time V , it is possible to express $\ K^{+}(x) $ in
terms of $\ I^{+}(x) $.

\vspace{5mm}

\noindent\textbf{Theorem 2:} \ If V is a globally hyperbolic $\
C^{0} $ - space - time, then
 $\ K^{+}(p) = cl ( int (K^{+}(p)), \\ p \in V $.

\noindent\textbf{Proof :} \ Let V be globally hyperbolic. It is
enough to prove that  $\ K^{+}(p) \subseteq \ cl ( int (K^{+}(p)),
\\ p \in \ V $. For this we show that $\ cl ( int (K^{+}(p)) $
is closed with respect to transitivity. So, let $\ x, y, z  \in \
cl ( int (K^{+}(p)) $ such that $\ x \prec \ y $ and $\ y \prec \
z $. We show that $\ x \prec \ z $. Since $\ x, y, z $ are limit
points of $\ int ( K^{+} (p))$, there are sequences $\ \{x_{n}\},
\  \{y_{n}\} , \ \{z_{n}\} $ in $\ int ( K^{+} (p) ) $ such that
$\ x_{n} \rightarrow \ x, \  y_{n} \rightarrow \ y, \  z_{n}
\rightarrow \ z $. Using first countability axiom , we may assume
, without loss of generality, that these sequences are linearly
ordered in the past directed sense ( see Martin and
Panangaden[16], lemma 4.3 ). Thus, for sufficiently large n, we
can assume that $\ x_{n} \prec \ y_{n}$ and $\ y_{n} \prec \ z_{n}
$. Since $\ x_{n}, \ y_{n}, \ z_{n} \in \ K^{+}(p) $, by
transitivity, $\ x_{n} \prec \ z_{n}$ for sufficiently large n. We
claim that $\ x \prec \ z $. Let $\ x $ be not in $\ K^{-}(z) $.
Then as $\ K^{-}(z) $ is closed,using local compactness, there
exists a compact neighbourhood N of x such that $\ N \cap K^{-}(z)
= \emptyset $, and so, $\ z $ is not in $\ K^{+}(N) $. Now, by
theorem 1, as V is globally hyperbolic and N is compact, $\
K^{+}(N) $ is closed. Hence,  there exists a K-convex
neighbourhood $\ N^{'} $ of z such that $\ N^{'} \cap K^{+}(N) =
\emptyset $, which is a contradiction as $\ x_{n} \prec z_{n}$ for
large n. Hence,  $\ x \prec \ z $. Thus, $\ cl ( int (K^{+}(p))$
is closed with respect to transitivity. Since, by definition, $\
K^{+}(p) $ is the smallest closed set which is transitive, we get,
$\ K^{+}(p) \subseteq \ cl ( int (K^{+}(p)) $. Hence  $\ K^{+}(p)
= \ cl ( int (K^{+}(p)) $. Similarly, $\ K^{-}(p) = \ cl ( int
(K^{-}(p)) $.

\vspace{5mm}

\noindent\textbf{Theorem 3:} \ If V is a globally hyperbolic $\
C^{0} $ - space - time then
 $\ int (K^{\pm}(x)) = I^{\pm}(x) , \ x \in V $.

\noindent\textbf{Proof :} \ Let V be globally hyperbolic and $\ x
\in \ V $. That $\ I^{+}(x) \subseteq \ int (K^{+}(x)) $ is
obvious by definition of $\ K ^{+}(x) $. To prove the reverse
inclusion, we prove that, if $\ x \prec \ y $ then there exists a
K-causal curve from $\ x $ to $\ y $ and if $\ y \in \ int(
K^{+}(x) )$, then this curve must be a future-directed time-like
curve.

Let $\ x \prec \ y $ and there is no K-causal curve from $\ x $ to
$\ y $. Then image of [0,1] will not be connected, compact or
linearly ordered. This is possible, only when a point or a set of
points has been removed from the compact set $\ K^{+}(x) \cap
K^{-}(y) $ , that is, when some of the limit points have been
removed from this set , which will imply that this set is not
closed.

But, since V is globally hyperbolic, $\ K^{+}(x) \cap K^{-}(y) $
is compact and hence closed. Hence, there must exist a K-causal
curve from $\ x $ to $\ y $.

Suppose, $\ y \in \ int (K^{+}(x)) $. Then, there exists a
neighbourhood $\ I^{+}(p) \cap I^{-}(q) $ of y such that  $\ y \in
\ I^{+}(p) \cap I^{-}(q) \subseteq K^{+}(x) $. To show that a
K-causal curve from $\ x $ to $\ y $ is time-like, it is enough to
prove that $\ x $ and $\ y $ are not null-related, that is, there
exists a non-empty open set in $\ K^{+}(x) \cap \ K^{-}(y)$.

Consider, $\ I^{+}(p) \cap I^{-}(q) \cap I^{+}(x) \cap I^{-}(y) $,
which is open. Take any point say z, on the future-directed
time-like curve from p to y . \\ Then, $\ z \in I^{+}(p) \cap
I^{-}(q) \cap I^{+}(x) \cap I^{-}(y)  \subseteq \ K^{+}(x) \cap \
K^{-}(y) $. ( Here, $\ z \in \ I ^{+}(x) $  because, if x and z
are null-related then $\ K^{+}(x) \cap \ K^{-}(z) $ will not
contain an open set. But $\ I^{+}(p) \cap \ I^{-}(z) \subseteq \
K^{+}(x) \cap \ K^{-}(z) $ ).  That is, $\ K^{+}(x) \cap \
K^{-}(y) $ has a non-empty open subset. Hence, x  and  y are not
null-related, and so, the K-causal curve from x  to  y is
time-like. That is, $\ y\in \ I^{+}(x) $. Thus, $\ int (K^{+}(x))
\subseteq I^{+}(x) $ which proves that $\ int (K^{+}(x)) = \
I^{+}(x) $. Similarly, we can prove that $\ int (K^{-}(x)) = \
I^{-}(x) $.

\vspace{5mm}

\noindent From theorems 2 and 3, we have the following result:

\noindent \emph{`For a globally hyperbolic space-time V , $\
K^{\pm}(x) = \ cl ( I^{\pm}(x)) $'}.

\vspace{5mm}

\noindent We now introduce the following condition given by Luca
Bombelli and \\ Johan Noldus [3] :- \\ \emph{`$\ K^{+}(p)
\subseteq K^{+}(q)$ and $\ K^{-}(q) \subseteq K^{-}(p) \Rightarrow
I^{+}(p) \subseteq I^{+}(q) $ and $\ I^{-}(q) \subseteq I^{-}(p)
$}'.

\noindent We call this as \emph{`Noldus condition'}.  Bombelli and
Noldus have given an interesting example of a space-time where
this condition is not satisfied.

\vspace{2mm}

\noindent We then have the following:

\vspace{5mm}

\noindent\textbf{Theorem 4:} \ The Noldus condition is equivalent
to $\ int (K^{\pm}(x)) = I^{\pm}(x) $ in a K-causal space - time.

\noindent\textbf{Proof :} \ Let V be a K-causal space - time and
$\ p, q \in \ V $. Let $\ int (K^{\pm}(x)) = I^{\pm}(x) $. Then,
$\ K^{+}(p) \subseteq K^{+}(q)$ and $\ K^{-}(q) \subseteq K^{-}(p)
\Rightarrow \ int (K^{+}(p)) \subseteq \ int (K^{+}(q)) $ and $\
int (K^{-}(q)) \subseteq \ int(K^{-}(p)) $. That is, $\ I^{+}(p)
\subseteq I^{+}(q) $ and $\ I^{-}(q) \subseteq I^{-}(p) $.

Conversely, let us assume the Noldus condition. Let $\ y \in \ int
( K^{+}(x) ) $. Then, there exists an open neighbourhood say, $\
I^{+}(p) \cap \ I^{-}(q) $ of y such that, $\ y \in I^{+}(p) \cap
\ I^{-}(q) \subseteq \ K^{+}(x) $. Therefore, $\ y \in I^{+}(p)$
and $\ q \in I^{+}(y) $.

Now, $\ y \in I^{+}(p)$ implies there exists a future directed
time-like curve from p to y. Choose a point z on this trip. Then,
$\ z \in I^{+}(p)$ and $\ y \in I^{+}(z) $ and hence $\  z \in \
I^{+}(p) \cap I^{-}(q) \subseteq \ K^{+}(x)$ . Thus  $\ x \prec \
z $. This then implies $\ K^{+}(z) \subseteq K^{+}(x)$ and also $\
K^{-}(x) \subseteq K^{-}(z) $ which implies $\ I^{+}(z) \subseteq
I^{+}(x) $ and $\ I^{-}(x) \subseteq I^{-}(z) $.  Thus, we have,
$\ y \in I^{+}(z) $ and $\ I^{+}(z) \subseteq \ I^{+}(x)$ which
gives $\ y \in \ I^{+}(x) $. Hence, $\ int (K^{+}(x)) \subseteq \
I^{+}(x) $ and so, $\  int (K^{+}(x)) = I^{+}(x) $.

\noindent Similarly, we can prove $\  int (K^{-}(x)) = I^{-}(x) $.

\vspace{5mm}

\noindent Hence from theorems 3 and 4 , it follows that global
hyperbolicity implies Noldus condition. We thus have,

\vspace{5mm}

\noindent\textbf{Theorem 5:} \ In a $\ C^{2} $ - space-time which
is K-causal, if Noldus condition is assumed then $\ J^{+} $ and $\
K^{+} $ are equal.

\noindent\textbf{Proof :} \ Let $\ x \in \ V $ and $\ y \in \
K^{+}(x) $. Then, $\ K^{+}(y) \subseteq K^{+}(x)$ and $\ K^{-}(x)
\subseteq K^{-}(y)$ and hence by assumption, $\  I^{+}(y)
\subseteq  \ I^{+}(x)$ and $\ I^{-}(x) \subseteq \ I^{-}(y)$.
Thus, $\ y \in \ J^{+}(x) $. Therefore, $\ K^{+}(x) \subseteq \
J^{+}(x) $ , and so, $\ K^{+}(x) = \ J^{+}(x) $.

\vspace{2mm}

\noindent Hence, we have the following:

\vspace{5mm}

\noindent\textbf{Corollary :} \ In a $\ C^{2} $ - space-time ,
Noldus condition implies $\ J^{+}(x) $ and $\ J^{-}(x) $ are
closed and hence such a  space-time is causally simple.

\vspace{5mm}

\noindent  From theorems  3, 4 ,5 and the above corollary, we have
the following:

\vspace{5mm}

\noindent\textbf{Corollary :} \ Noldus condition lies in between
global hyperbolicity and causal simplicity.

\vspace{8mm}

\section{K- causality hierarchy and Concluding remarks:} \textbf{(i)}
In this paper we have generalized concepts from causal structure
theory in terms of K-causal relations and causal maps in the light
of the work of Sorkin and Woolgar [26] and Garcia-Parrado and
Senovilla [8 , 9]. We have proved a number of results in this
context as it is seen from the text of the paper.

In section 3 , we have proved that strong causality with respect
to $\ K^{+} $ implies K-future distinguishing. Thus K-causality
implies strongly causality with respect to K  which implies
K-distinguishing. Since  a K-causal space-time is always
K-reflecting as remarked in section 3, it follows that the
K-causal space-time is K-reflecting as well as K-distinguishing.
In the classical theory, such a space-time is called causally
continuous[12]. Such space-times have been useful particularly in
the study of topology change in quantum gravity [7]. Thus if we
define a K-causally continuous space-time analogously,  then we
get the result that a K- causal $\ C^0 $ space-time is K- causally
continuous. Moreover, since $\  K^{\pm} (x)$ are topologically
closed by definition, analogue of causal simplicity is redundant.
Since causal continuity implies stable causality in the classical
sense[12], we expect that K-causal space-time is stably causal. As
remarked in [26], stable causality implies K- causality. Thus we
shall then have equivalence of these two conditions. Recently,
certain steps towards this equivalence have been proved by
Minguzzi [18], though complete equivalence is yet to be proved.
Assuming this equivalence,\emph{ K-causal hierarchy }would read as
follows:

\vspace{2mm}

\textbf{Global hyperbolicity \ $\ \Rightarrow $ \  Stably causal \
$\ \Leftrightarrow $ \  K- causality \ $\ \Rightarrow $ \ Strong
causality \ $\ \Rightarrow $ \ Distinguishing}

\vspace{2mm}

The recent results proved by Minguzzi [18] and earlier results
proved by Dowker, Garcia and Surya  [6] support and complement the
results proved in section 3 and also this hierarchy. In [6] , the
definition of causal continuity has been kept intact as in the
classical sense, and hence the results differ from those proved
here. However, [6] discusses in addition , the role of $\  K^{+}$
in topology changing Morse space-times both with and without
degeneracies and find further characterizations of causal
continuity. Thus they prove that a Morse space-time is causally
continuous if and only if the functions Int($\  K^{+}$(.)) and
Int($\  K^{-}$(.)) are inner continuous. On the other hand, our
paper discusses K- causal maps and their properties as an
additional feature, as compared to the papers [6] and [18].

\vspace{2mm}

\textbf{(ii)}. Martin and Panangaden[16, 17] have defined a
bicontinuous poset  $\ (X,\leq )$  as \emph{globally hyperbolic},
if the intervals [a,b] are compact in the interval topology.
Similar definition has been given, using causal intervals in the
theory of convex cones in the book \emph{Causal symmetric spaces}
[13]. In Martin and Panangaden[16] ( see theorem 4.1), it is
proved that in a globally hyperbolic poset, its partial order $\
\leq $ is a
 closed subset of $\ X \times X $.

In a K-causal space - time, since the relation we have studied in
this paper is a closed partial order, and global hyperbolicity is
defined in terms of this relation, it is natural to ask whether
the results in Martin and Panangaden [16, 17] can be proved for
 K-causal space - time. Work is in progress in this direction.

 \vspace{10mm}

\noindent\textbf{Acknowledgements :} \ We would like to thank
Professors E. Woolgar, R.D. Sorkin, \\ A. Garcia-Parrado and J.
Senovilla for many clarifications and encouragement during the
course of this work. We also thank Dr. S. H. Ghate for the
interest he has taken in this work and for his valuable
suggestions.

\newpage

\section{References}

\hspace{4mm} \vspace{2mm} [1] \ J\ K\  Beem, \emph{General
Relativity and Gravitation}, \textbf{8} , No.4,  245 (1977).

\vspace{2mm} [2] \ J\ K\ Beem, P\ E\  Ehrlich and K\ L\ Easley,
\emph{Global Lorentzion Geometry},  (Monographs textbooks. Pure
Appl Mathematics, Dekker Inc., New York 1996).

\vspace{2mm} [3] \ L\ Bombelli and J\ Noldus, \emph{Class.Quantum
Grav.},
 \textbf{21}, 4429 (2004).

\vspace{2mm} [4] \ C\ J\ S\ Clarke  and P\ S\ Joshi,
\emph{Class.Quantum Grav.}, \textbf{5}, 19-25 (1988).

\vspace{2mm} [5] \ J\ Dieckmann, \emph{Gen. Rel. Grav.}
\textbf{20} ,  No.9, 59 (1988).

\vspace{2mm} [6] \ H\ F\ Dowker, R\ S\ Garcia, S\ Surya,
\emph{Class. Quantum Grav.,} \textbf{17},  4377 - 4396 (2000).

\vspace{2mm} [7] \ H\ F\ Dowker, R\ S\ Garcia, S\ Surya,
\emph{Class. Quantum Grav.,} \textbf{17,}  697 - 712 (2000).

\vspace{2mm} [8] \ A\ Garcia-Parrado and J\ M\ Senovilla,
\emph{Class.Quantum Grav.} \textbf{20} , 625 (2003).

\vspace{2mm} [9] \ A\ Garcia-Parrado and J\ M\  Senovila,
\emph{Class.Quant.Grav.}, \textbf{21},  661-696, (2004).

\vspace{2mm} [10] \ R\ Geroch, \emph{J Math Phys,}  \textbf{11} ,
437 (1970).

\vspace{2mm} [11] \  S\ W\ Hawking and G\ F\ R\ Ellis,  \emph{The
Large Scale Structure of Space-time}, (Cambridge Univ Press,
1973).

\vspace{2mm} [12] \ S\ W\ Hawking and R\ K\  Sachs,
 \emph{Commun. Math. Phys.}, \textbf{35}, 287 (1974).

\vspace{2mm} [13] \ J\ Hilgert and G\ Olafsson,  \emph{Causal
symmetric spaces,
 Geometry and Harmonic Analysis}, ( Academic Press, New York  1997).

\vspace{2mm} [14] \ P\ S\ Joshi,  \emph{Global Aspects in
Gravitation and Cosmology}, ( Oxford Science Publications, 1993).

\vspace{2mm} [15] \ K\  Martin, \emph{Class.Quant.Grav.},
\textbf{23}, 1241-1252 (2006).

\vspace{2mm} [16] \ K\  Martin and P\ Panangaden,
\emph{gr-qc}/0407093.

\vspace{2mm} [17] \ K\  Martin and P\ Panangaden,
\emph{Commun.Math.Phys}., \textbf{267}, 563-586  (2006).

\vspace{2mm} [18] \ E\  Minguzzi,  \emph{gr-qc}/0703128.

\vspace{2mm} [19] \ R\ Penrose,  \emph{Techniques of Differential
Topology in Relativity}, ( AMS Colloquium Publications,  1972).

\vspace{2mm} [20] \ R\ Penrose, R\ D\ Sorkin and E\ Woolgar,
\emph{gr-qc}/9301015.

\vspace{2mm} [21] \ M\ Rainer,  \emph{Int. J. Mod. Phys.},
\textbf{ D 4}, 397-416 (1995).

\vspace{2mm} [22] \ M\ Rainer,  \emph{Gravitation and Cosmology},
\textbf{ 1}, 121-130 (1995).

\vspace{2mm} [23] \ D\ P\ Rideout   and  R\ D\ Sorkin, \emph{Phys.
Rev.}, \textbf{D 61}, 024002 (2000).

\vspace{2mm} [24] \ R\ D\ Sorkin \emph{gr-qc}/0309009.

\vspace{2mm} [25]  \ R\ D\ Sorkin \emph{gr-qc}/ 0710.1675.

\vspace{2mm} [26] \ R\ D\  Sorkin and E\ Woolgar,
\emph{Class.Quantum Grav.}, \textbf{3},  1971 (1996).

\vspace{2mm} [27] \ R\ Wald , \emph{General Relativity}, (Univ of
Chicago Press, 1984).

\vspace{2mm} [28] \ E\ C\ Zeeman, \emph{J. Math. Phys,}
\textbf{5}, 490 (1964).

\end{document}